\documentclass[prl,preprint,,amsmath]{revtex4}
\input {epsf.sty}
\usepackage{epsfig}
\usepackage{amssymb}
\begin{document}
\title{The Stokes-Einstein Relation in Supercooled Aqueous Solutions
of Glycerol}
\author{Bo Chen, E.E. Sigmund\footnote{present address:
Schlumberger-Doll Research, Ridgefield, CT 06877}, and W.
P. Halperin}
\affiliation{Department of Physics and Astronomy,\\
      Northwestern University, Evanston, Illinois 60208}

\date{Version \today}

\begin{abstract} The diffusion of glycerol molecules decreases with
decreasing temperature as
its viscosity increases in a manner simply described by the
Stokes-Einstein(SE) relation.  Approaching the glass
transition, this relation breaks down as it does with a number of
other pure liquid glass formers.
We have measured the diffusion coefficient for binary mixtures of
glycerol and water and find that the
Stokes-Einstein relation is restored with increasing water concentration.
Our comparison with theory suggests that addition of water postpones 
the formation of
frustration domains\cite{Kivelson}.
\end{abstract}

\maketitle

\vspace{11pt}

Most liquids become solids after they are cooled through their
melting temperature,
$T_m$, following a first order thermodynamic transition. A notable
exception is a class of
molecular fluids known as glass formers which can stay in a
metastable state as supercooled
liquids to temperatures substantially below their equilibrium melting
transition.  The thermodynamic
properties of the supercooled liquid are  a continuous extrapolation
of the liquid state from above
$T_m$.   On further cooling the dynamical and transport behavior,
viscosity, diffusion,
and motional relaxation times, exhibit such dramatic slow-down over a
narrow temperature range, that
there appears to be a transition to a rigid but amorphous solid state
at a  characteristic temperature, $T_g$,
called the glass transition\cite{Deb01}.

According to the Stokes-Einstein (SE) relation the product of viscosity
and diffusion coefficients is simply proportional to the absolute
temperature\cite{Einstein}.

\begin{eqnarray}
D=\frac{k{_B}T}{6\pi{\eta}r_s},
\end{eqnarray}

In Eq.1, $D$ is the translational diffusion
coefficient, $\eta$  is the viscosity,
   $k{_B}$, the Boltzmann constant, and $r{_s}$ the hydrodynamic
radius. The basis for the SE-relation is that the elementary units of the
liquid are just
Brownian particles and that the system is homogeneous. It is remarkable
that many glass formers still follow this behavior even as the
relaxation processes slow-down. Ultimately,
at low enough temperature, it fails, at least for a number of the
more fragile\cite{Ang88} glass
forming liquids.

Kivelson\cite{Kivelson} {\it et al.} have developed a theory for 
frustation limited domains (FLD)
that form at a temperature T* above the melting temperature $T_m$.
Although these domains are energetically favorable on a  local scale 
their existence  affects
diffusional and rotational relaxation differently giving rise to 
breakdown of the SE-relation. In this
letter we present diffusion measurements  of four aqueous solutions 
of glycerol and compare with
viscosity measurements to show that the SE-relation fails for 
glycerol; but  that this relation can be
restored at low temperatures through addition of water. With the FLD 
theory we extract the
temperatures for domain formation,  T*,  for the four solutions
and we show they are consistent
with the  temperatures where we first observe deviation from the SE-relation,
  providing a direct confirmation of these theoretical ideas.  We have 
chosen binary solutions
to study glass formation since they allow systematic and continuous 
control of the essential
molecular interactions involved, an approach that may also prove 
useful for numerical
simulations.

Among glass formers, glycerol is well studied having an
intermediate fragility\cite{Plazek}
m = 53, glass transition temperature
$T_g$ = 190 K and melting temperature $T_m$ = 290 K. It is strongly
resistant to
crystallization\cite{Boutron,Cocks} as are its aqueous
solutions\cite{Z. Chang, C.
Gao,SSNMurthy}. For glycerol concentrations above 55\% they are good
single-phase glass
formers\cite{Huck}. The concentration dependent glass transition
temperature has been
measured\cite{Rasmussen,Murthy} and can be described quite well by an
empirical formula\cite{Jenckel}:
$T_g=190w_1+136w_2+27w_1w_2$,
where $w_1$ and $w_2$ denote the weight concentration of glycerol and water
respectively, such that $T_g$ smoothly decreases with increasing
water content as is shown in Fig.2.

Samples were prepared with 100, 95, 90 and 85\%  glycerol by weight,
using pure glycerol (99.9\%, Fisher
Scientific) and D$_2$O (99.9\%, Cambridge Isotopes).
The $^1$H NMR signal, which we used for measurement of diffusion, was
therefore selectively sensitive to the
diffusion of the glycerol molecule. The mixtures were loaded into
pyrex glass ampoules in the dry nitrogen atmosphere of a glove bag.
The ampoules were flame-sealed while the liquid mixture inside was
frozen with liquid nitrogen.  The
diffusion measurements were performed by proton NMR stimulated-echo
sequences in the fringe magnetic field, H =
4.28 T, $^1$H frequency of 182 MHz, of a superconductive magnet with
a magnetic field gradient G = 42.3
T/m. In order to improve signal-to-noise, and hence resolution in
diffusion, we used the frequency jumping method
developed by Sigmund and Halperin\cite{Sig03,Eric}, allowing the
near-simultaneous collection of data from 17
independent NMR-sample slices, each 110 microns thick.

The slow-down in dynamics of a glass former is marked by deviation
of the temperature dependence of diffusion
from the Arrhenius behavior that is typical at high temperatures:
$\propto  e^{-E/k_{B}T}$. The modified
temperature dependence can usually be represented by the
Vogel-Tamman-Fulcher (VTF)
equation\cite{Vogel,Tamman,Fulcher} which, for the diffusion
coefficient D, is given by,
\begin{eqnarray} D = A_0\sqrt{T}exp(-\frac{B}{T-T_0}).
\end{eqnarray}
Here $A_0$, $B$, and $T_0$ are phenomenological constants determined
from experiment. At still lower
temperatures, close to
$T_g$, the non-exponential features of relaxation, diffusion, and
viscosity are often best
captured by stretched exponential behavior. A theoretical
basis that describes the
continuous evolution of glass forming dynamics with decreasing
temperature has not yet been established. Nonetheless, it is clear 
that the molecular
interactions must play a  crucial role.

\begin{figure}[h]
\vspace{0.1in}
\includegraphics[width=8 cm]{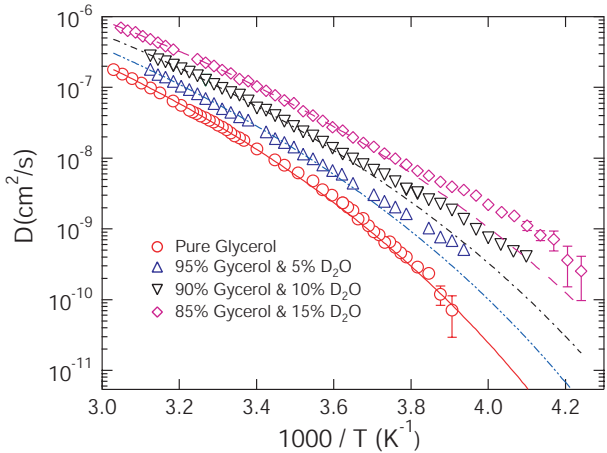}
\suppressfloats [b]
\caption[Diffusion of glycerol]{The logarithm of the diffusion
coefficient D of glycerol as a function of inverse temperature for 
various aqueous
solutions of glycerol. The curves are  fits to the VTF-formula. }
\label{Fig1}
\end{figure}

The logarithm of the diffusion coefficient for glycerol is plotted as a
function of the inverse temperature in Fig.1 to emphasize deviation
from  Arrhenius behavior which would be a straight
line in this figure. The diffusion coefficients shift to the right
with increasing water content while
retaining a temperature dependence similar to that for pure glycerol,
at least in the high temperature region.
But at lower temperatures the diffusion
for solutions  is softened compared
to that of pure glycerol. To get a quantitative assessment, we fit
the experimental diffusion coefficients to the
VTF-relation and present the important parameters for these fits in
Fig.2 along with T*, determined from the viscosity
data\cite{Segur, Schroter} analyzed using the frustration limited 
domain theory (FLD)
\cite{Kivelson}, and $r{_s}$, the hydrodynamic radius obtained from 
the SE-relation,
Eq.1.

The VTF-relation, Eq.2,  can be derived from a free volume
theory\cite{Morrel}. The prefactor
$A_0$ in this relation is a constant, at most a weak function of
water concentration; the activation energy, B,
was found to be a weak, but systematically decreasing, function of
water concentration; and the temperature,
$T_0$, corresponds to full arrest of the molecular dynamics
at a temperature where there would be zero free volume.  This
temperature should be
less than, but closely related to, the glass transition
temperature\cite{Deb01}. A direct comparison between
the two is given in Fig.2, showing that $T_0$ scales with $T_g$.

\begin{figure}[h]
\vspace{0.1in}
\includegraphics[width=8 cm]{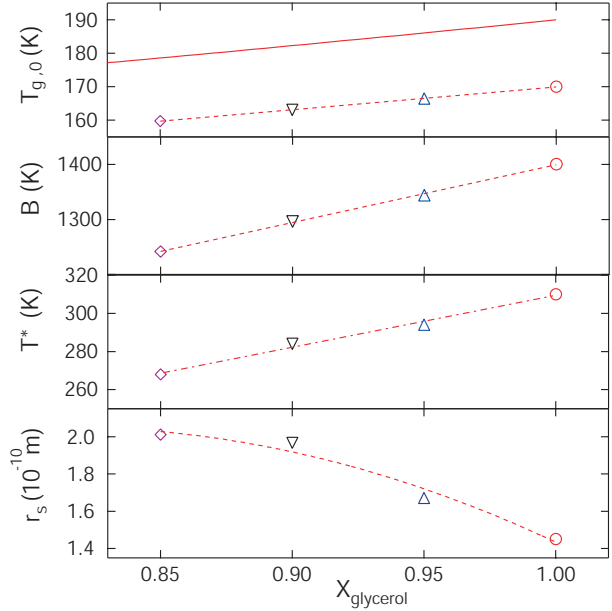}
\suppressfloats [b]
\caption[VTF-parameters and r_s]{VTF-parameters, the domain
formation temperature, T*,  and the hydrodynamic radius,
$r_s$, as a function of glycerol
concentration in aqueous solutions. The solid curve is  the glass 
transition temperature
$T_g$, taken from the empirical formula given in the text.  It 
closely parallels
  the VTF-parameter
$T_0$. The dashed lines are guides to the eye.}
\label{Fig2}
\end{figure}

The VTF-fits shown in Fig.1 were performed for diffusion
coefficients in the high temperature region. They
deviate from the  experimental values at low temperature where  $D
\approx 10^{-8}$ cm$^{2}$s$^{-1}$. In order to understand the
failure of the VTF-formula at low temperature, we return to the free
volume theory
from which the VTF-relation is derived. In this theory, the diffusion of the
molecule is associated with the potential cage formed by its
neighbors\cite{Morrel}, such that the molecular
interactions are described in a mean-field way. In the case of
glycerol and its aqueous solutions, the dominant
molecular interaction is hydrogen bonding (H-bonding) and the
potential cage should arise from these bonds.
Numerous theoretical investigations on H-bonding have been conducted
for glycerol, water and other
H-bonding dominated alcohols\cite{Chelli, Padro, Marti, Matsumoto}
and it has been shown that the detail of
H-bonding and its effects are complex: one molecule can have
different H-bonding states depending on the number of
bonds; and H-bonding can be distinguished as either intermolecular or
intramolecular\cite{Chelli}, affecting diffusion differently.  The life
time of the H-bond is also
temperature dependent\cite{Padro, Chelli}.
The more intermolecular H-bonding a molecule has, the more of
an obstacle there is to
diffusion. This intuitive idea is confirmed in calculations by Marti {\it
et al.}\cite{Marti} for water and by Matsumoto {\it et
al.}\cite{Matsumoto} for methanol. The molecular
dynamics  calculations by Chelli {\it et al.}\cite{Chelli} and Padro
{\it et al.}\cite{Padro} indicate
that the average number of H-bonds per molecule is higher for
glycerol than water, and their life time in
glycerol is about 6 times longer\cite{Padro}. The average life time of a
H-bond in glycerol
can be  described by an Arrhenius
equation\cite{Chelli} but the state of H-bonding also varies with
temperature\cite{Chelli}.  This complicated
situation goes beyond the simple free volume picture\cite{Morrel} and
leads to failure of the VTF-relation at low
temperature where these additional temperature dependences come into play.

Fits to the SE-relation in the high temperature limit give the
results for $r_s$ in Fig.2 with a satisfying
consistency for different water mixtures but systematically
increasing by 25\% over the range of water
concentrations.  In terms of H-bonding we argue that a
dynamical complex of glycerol and water
in binary mixtures is the relevant elementary unit, akin to the ideal
Brownian particle, having a larger effective
radius than pure glycerol which increases systematically with water
content.  The increase in the diffusion coefficient of glycerol with 
dilution in aqueous
mixtures can be
ascribed to the decrease in the effective interaction between these
Brownian clusters that we associate with the weaker intermolecular
H-bonding in water.

  In Fig.3 we show the temperature dependence relative to the glass 
transition of $D{\eta}/T$,
normalized at high temperature.
The viscosity data
for pure glycerol are  taken from smooth fits of two data
sets\cite{Segur,Schroter} which are consistent with each other to 
$\approx 3\%$.
The viscosity of the aqueous solutions are taken from smooth
fits to the  data by Segur {\it et. al.}\cite{Segur}.
At temperatures $T/T{_g} <
1.62$ we resolve a breakdown in the
SE-relation for pure glycerol beyond the estimated measurement error
in diffusion which is represented by the scatter in
our data.  The temperature at which this breakdown is observed moves
to lower temperature with increasing amounts of water.
In fact, for the 85\% sample the SE-relation holds over the entire
range for which there are viscosity data available.   The behavior of pure
glycerol we report here matches qualitatively that of
other glass formers  where larger deviations from the SE-relation are
observed at temperatures closer to their glass transition.
Collectively, these results on glass formers are consistent
with theoretical models that have recently been proposed\cite{Jun04}
suggesting failure in the SE-relation irrespective of fragility.

\begin{figure}[h]
\vspace{0.1in}
\includegraphics[width=8cm]{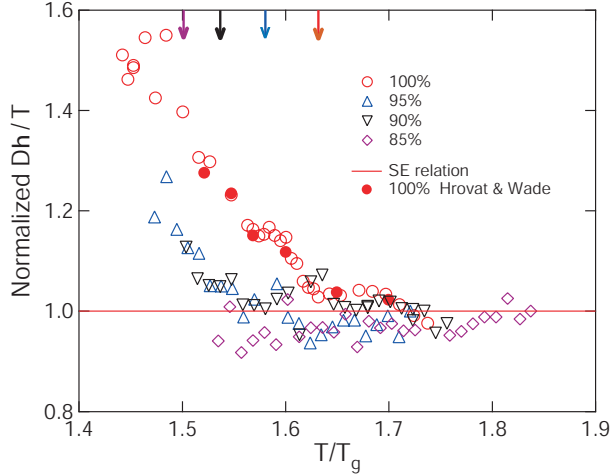}
\caption[Stokes-Einstein comparison versus inverse
temperature]{Breakdown of the Stokes-Einstein relation
through comparison of diffusion and  a smoothed interpolation to
viscosity data\cite{Segur,Schroter} for aqueous solutions
of glycerol.  For comparison the diffusion data
of Hrovat and Wade are also shown for pure
glycerol\cite{Hro81} with the same normalization. The validity of the
Stokes-Einstein relation is limited to high
temperatures. The lower limit
of temperature in this figure is determined by availability of
viscosity data in aqueous mixtures. The four arrows on the top axis
indicate T*, the domain formation temperature. From left to
right they are for 85, 90, 95 and 100\% glycerol solutions, and are
qualitatively consistent
with onset for breakdown of the SE-relation.}
\label{Fig3}
\end{figure}

For pure liquids, such as glycerol,
Arrhenius behavior is observed at high temperatures in both diffusion
and viscosity.  So it is natural to
expect the SE-relation to hold there.  But at lower temperature,
where VTF-behavior is observed,  it is not obvious
that the SE-relation should remain valid; or if not, what  might be
the mechanism responsible for its failure.
One important aspect of this problem is that the viscous relaxation process is
   microscopic while the measurement of diffusion is macroscopic,
meaning that the former is on the molecular length
scale and the latter on the length scale of tens of microns.  So a possible
explanation for breakdown of the SE-relation could be  ascribed to
local domain or cluster formation which can lead
to dynamical heterogeneity\cite{Kivelson,MTCicerone,Cic95,Colby,Jun04}.

Kivelson {\it et al.}\cite{Kivelson} suggested that the breakdown of
the SE-relation is due to dynamical heterogeneity,  resulting from
formation and
growth of frustration limited domains
below a temperature T*.  We have analyzed the viscosity
data\cite{Segur,Schroter} using the FLD
theory\cite{Kivelson} to find T* plotted in Fig.2, and given in Fig.3 by
vertical arrows. The temperature where deviation
from the SE-relation appears in Fig.3, is qualitatively consistent
with the domain formation temperature T*.  In particular they both decrease
systematically with increasing water content,
suggesting that the breakdown of the SE-relation is controlled by the
birth of these frustrated domains.

It is now well-established experimentally that the SE-relation fails for
certain fragile glass formers in their
pure form such as ortho-terphenyl\cite{Fuj92} and
tris-napthylbenzene\cite{Swa03}, and now for glycerol, a 
significantly less fragile material.
However, Chang
and Sillescu\cite{Cha97} did not find such breakdown behavior for  glycerol.
This is in contrast to our
preliminary work\cite{Eric} and that which we report here. There are 
significant
differences ($\approx 30\%$)
in the absolute values of the diffusion coefficients for pure 
glycerol measured by Chang and
Sillescu\cite{Cha97} as
compared with other workers\cite{Hro81,Tom73} including the
more precise measurements of Hrovat and Wade\cite{Hro81} which
we have confirmed and extended to lower temperatures; see Fig.3.
This discrepancy is outside
the possible error in our measurements of
diffusion and it is most likely that there is systematic error in the Chang and
Sillescu data  that accounts for their
claim that glycerol obeys the SE-relation over this temperature range.
We also note that the work of these authors is
inconsistent with the FLD theory.

In conclusion, we have measured the diffusion coefficient of glycerol
in aqueous solution.  We have found that
VTF-behavior fails as a phenomenological form for the temperature
dependence of the diffusion coefficient, but
with a range of validity that extends to lower temperature with
increasing water content.  Similarily, we find
that the SE-relation breaks down for these aqueous
mixtures but at a temperature that decreases
progressively with increasing water content.  These observations
are consistent with expectations from
frustration limited domain theory\cite{Kivelson,Jun04}. Finally, we
remark that the  mechanism responsible
is most likely the heteromolecular hydrogen bonding which postpones the
formation of domains and inhibits the breakdown of the Stokes-Einstein
relation as temperature is decreased. We find that the effective intermolecular
interaction from hydrogen bonding can be conveniently controlled 
using binary solutions.

We thank M.D. Ediger for helpful suggestions and acknowledge support
from the Materials Research Center at
Nortwestern University provided by the NSF through grant DMR-0076097
and support from the Department of Energy,
contract DE-FG02-05ER46248.

\end{document}